\documentclass[conference]{IEEEtran}
\IEEEoverridecommandlockouts
\usepackage{cite}
\usepackage{amsmath,amssymb,amsfonts}
\usepackage{algorithm}
\usepackage{algpseudocode}
\usepackage{graphicx}
\usepackage{subfigure}
\usepackage{textcomp}
\usepackage{xcolor}

\def\BibTeX{{\rm B\kern-.05em{\sc i\kern-.025em b}\kern-.08em
    T\kern-.1667em\lower.7ex\hbox{E}\kern-.125emX}}

\ifodd 1
\newcommand{\congr}[1]{{\color{blue}#1}}
\else
\newcommand{\congr}[1]{#}
\fi

\ifodd 1
\newcommand{\congc}[1]{{\color{red}(Cong: #1)}}
\else
\newcommand{\congc}[1]{}
\fi

\begin{document}

\title{Average Reward Reinforcement Learning for Wireless Radio Resource Management
\thanks{The work of K. Yang and C. Shen was partially supported by the U.S. National Science Foundation (NSF) under awards CNS-2002902, CNS-2003131, ECCS-2029978, ECCS-2030026, ECCS-2143559, and SII-2132700. The work of J. Yang was supported in part by the U.S. NSF under awards CNS-1956276, CNS-2003131 and CNS-2030026.}
}
% \thanks{The work of K. Yang and C. Shen was partially supported by the U.S. National Science Foundation (NSF) under awards CNS-2002902, CNS-2003131, ECCS-2029978, ECCS-2030026, ECCS-2143559, and SII-2132700. The work of J. Yang was supported in part by the U.S. NSF under awards CNS-1956276, CNS-2003131 and CNS-2030026.}

\author{\IEEEauthorblockN{Kun Yang$^*$, Jing Yang$^\dag$, and Cong Shen$^*$}%, Shu-ping Yeh$^\ddag$, Jerry Sydir$^\ddag$}
\IEEEauthorblockA{$^*$ Department of Electrical and Computer Engineering, University of Virginia, USA\\
$^\dag$ Department of Electrical Engineering, The Pennsylvania State University, USA\\
%$^\ddag$ Intel Corporation, USA
}}

\maketitle

\begin{abstract}
In this paper, we address a crucial but often overlooked issue in applying reinforcement learning (RL) to radio resource management (RRM) in wireless communications: the mismatch between the \emph{discounted} reward RL formulation and the \emph{undiscounted} goal of wireless network optimization. To the best of our knowledge, we are the first to systematically investigate this discrepancy, starting with a discussion of the problem formulation followed by simulations that quantify the extent of the gap. To bridge this gap, we introduce the use of \emph{average reward} RL, a method that aligns more closely with the long-term objectives of RRM. We propose a new method called the Average Reward Off-policy Soft Actor-Critic (ARO-SAC), which is an adaptation of the well-known Soft Actor-Critic algorithm in the average reward framework. This new method achieves significant performance improvement -- our simulation results demonstrate a $15\%$ gain in the system performance over the traditional discounted reward RL approach, underscoring the potential of average reward RL in enhancing the efficiency and effectiveness of wireless network optimization. 
\end{abstract}

\begin{IEEEkeywords}
Radio resource management, averaged reward reinforcement learning, deep reinforcement learning
\end{IEEEkeywords}

\section{Introduction} \label{sec:intro}

In recent years, there has been a growing interest in applying reinforcement learning (RL) methods to solving radio resource management (RRM) problems in wireless networks. It largely stems from several important RL properties that match the characteristics of wireless networking. First, wireless network optimization is a {closed-loop} and {sequential} operation: set parameters, observe performance, and fine-tune. Second, many tasks in wireless networks have \emph{long-term} performance impact, and their parameters are adjusted at a very low pace. As a result, the optimization cannot only target the {immediate} performance gain, but must take a long-term view. Third, there exist well-established {feedback protocols} in wireless standards, which provide a built-in mechanism for observing the state and receiving rewards. Lastly, RL research is a highly active and theoretically well-grounded area of machine learning, which lays a good foundation to its success in wireless networking.

Despite the promising initial results and the philosophical match, the majority of the existing solutions rely on the standard RL formulation, which maximizes \emph{discounted} cumulative rewards in the long term. This objective, however, is misaligned with the typical objectives of wireless network optimization, where we do not treat future utility less importantly than the current one. A typical example is that we generally try to maximize the long-term average throughput of the entire network, treating both current and future user throughput equally in this formulation. %For such problems, the existing RL solutions based on discounted reward maximization suffer

Naturally, one would ask whether we can design RL solutions for wireless network optimization that directly use undiscounted total reward as the objective. In the RL literature, this falls into the category of \emph{average reward RL} \cite{barto2021reinforcement}. To the best of the authors' knowledge, such average reward-based RL solutions have not been developed in wireless network optimization. In fact, the field of average reward RL itself is relatively under-explored. Only until recently have we seen the advancements to extend Policy Proximal Optimization (PPO) \cite{zhang2021average} and Deep Deterministic Policy Gradient (DDPG) \cite{saxena2023off} into the average reward framework. Nevertheless, these developments signal a growing potential for applying average reward RL in real-world engineering applications.

% The computational power boost enabled deep learning to play a more important role in various areas. In particular, for sequential decision-making tasks such as radio resource management (RRM) in wireless communication systems, reinforcement learning (RL) has become increasingly vital. Despite its popularity, the standard discounted reward RL formulation is misaligned with the objectives of wireless optimization, highlighting an opportunity for advancement.

% Some of the previous research in discounted reward RL has observed that for problems characterized by longer horizons and stable states, a higher discount factor $\gamma$ tends to improve policy performance \cite{tarasov2024revisiting, wu2022supported}. This observation suggests these problems could benefit from a setting considering outcomes from a longer future. 

% On the other hand, the average reward RL \cite{barto2021reinforcement}, which has an aligned objective with the RRM problem in wireless communication systems, remains underutilized due to a lack of practical deep-learning applications. Nevertheless, recent advancements have integrated techniques such as Policy Proximal Optimization (PPO) \cite{zhang2021average} and Deep Deterministic Policy Gradient (DDPG) \cite{saxena2023off} into the average reward framework. These developments signal a growing potential for applying average reward RL in practical scenarios.

In this paper, we begin by pinpointing the discrepancy between the widely used discounted reward RL and the commonly adopted goals that are specific to RRM problems in wireless networks. Subsequently, we cast the RRM problem in an average reward RL framework. We develop a novel extension of the popular Soft Actor-Critic (SAC) algorithm to the average reward RL formulation, enhancing its applicability and effectiveness in addressing the RRM challenge. Our main contributions are summarized as follows.
\begin{enumerate}
    \item To the best of our knowledge, we are the first to identify the discrepancy between the \emph{discounted} reward RL formulation and the \emph{undiscounted} objective of wireless network optimization. We showcase this discrepancy by re-formulating a RAN network slicing problem as an averaged reward RL one, and highlighting the mismatch of the design objectives of the prior RL approaches. We achieve this by unequivocally demonstrating the impact of the discount factor $\gamma$ and environmental horizon on RAN slicing RRM in the prior solutions via numerical experiments. %, revealing that extending the horizon and increasing the discount factor can enhance system performance even under a discounted reward framework. 
    % \item Through experimental validation, we demonstrate the influence of the discount factor $\gamma$ and environmental horizon on RAN slicing RRM, revealing that extending the horizon and increasing the discount factor can enhance system performance even under a discounted reward framework. 
    \item {We investigate how the practical algorithms handle the challenges of average reward rate estimation, and how the RL update is performed. Based on the estimation strategy for the off-policy RL algorithms introduced in ARO-DDPG \cite{saxena2023off}, we extend the popular off-policy deep RL algorithm SAC to an average reward version called \textbf{ARO-SAC} (Average Reward Off-policy SAC). With a tweak to the conventional TD error and Bellman equation, our new design enables SAC to perform with the average reward objective.} 
    % propose Average Reward Off-Policy SAC (ARO-SAC), an empirical extension following the work ARO-DDPG \cite{saxena2023off}. \congc{Right now you only have one sentence, which is insufficient. Add more sentences here to highlight your contribution to developing this algorithm. }
    \item Our experimental result using an industry-grade wireless network simulator reveals that, with a properly selected hyperparameter, the proposed ARO-SAC can outperform the best SAC by a performance gain of $15\%$. We further investigate how the learning rates for the average reward rate and the environment horizon impact the performance of ARO-SAC. %Showing the hyperparameter's impact on ARO-SAC performance under our RRM problem.}
    
    % Our findings indicate that with proper hyperparameter optimization, the proposed Average Reward SAC surpasses its discounted counterpart by a margin of $15\%$.\congc{same as above -- you need to add more sentences to highlight your contributions here.}
\end{enumerate}

The rest of the paper is organized as follows. The related works are surveyed in Section \ref{sec:related}. In Section \ref{sec:setting}, we formulate the RRM problem using discounted reward RL and discuss the objective mismatch. We investigate the impact of discount factor and horizon in Section \ref{sec:understand}, which motivates us to develop the ARO-SAC algorithm in Section \ref{sec:average}. Section \ref{sec:con} concludes the paper.

% In Section \ref{sec:related}, we introduce related works, including discussions on deep reinforcement learning for radio resource management and averaged reward reinforcement learning. Section \ref{sec:setting} outlines the problem setting, highlighting the misalignment between traditional system goals and those of discounted RL. We then initiate our exploration into enhancing the performance of the discounted reward RL and understanding the gap in Section \ref{sec:understand}, demonstrating improved outcomes with larger discount factors and extended horizons. Section \ref{sec:average} proposes average reward RL as the solution to bridging the aforementioned gap, discussing the differences between average reward RL and discounted reward RL, and empirically extends the Soft Actor-Critic (SAC) algorithm to an averaged reward version, building upon the previously introduced average Deep Deterministic Policy Gradient (DDPG).  The paper concludes with Section \ref{sec:con}, where we discuss potential future research directions. \congc{This is way too long. You have done this multiple times in your prior papers -- no need to elaborate on every detail in the outline. Just say what each section is about. Each section should only use one short sentence.}

\section{Related Works} \label{sec:related}

\textbf{RL for RRM:} Due to its natural fit, RL-based solutions have been gradually adopted to solve RRM problems. Previous efforts include the solutions based on the bandit algorithms \cite{nagaraja2016power,valliappan2016base, huang2016methods, Shen2018jstsp, Zhou2019twc}. Subsequently, Q-learning-based algorithms were developed \cite{ahmed2019deep,meng2019power, zhao2019joint}, followed by the adoption of the actor-critic architecture \cite{nasir2020deep, Yang2020infocomwksp, polese2022coloran, foukas2016flexran, nasir2021deep}. Regarding decentralized methods, multi-agent reinforcement learning (MARL) has solidified its relevance in \cite{nasir2019multi, yang2022multi, naderializadeh2021resource, zhang2023distributed}. More recent research has explored training RL policies using offline datasets \cite{yang2024offline, yang2024advancing}. Despite these advancements, all methods predominantly rely on \textit{discounted reward} RL algorithms, which overlooks a crucial aspect: the mismatch between the traditional objectives of wireless systems and the principles underlying discounted reward RL. %This paper proposes a paradigm shift towards average reward RL, which better aligns with wireless systems' long-term objectives, potentially enhancing resource management strategies' effectiveness and sustainability.

\textbf{Averaged reward RL:} Average reward RL, as a different formulation from the discounted reward RL setting, was designed to handle the scenarios where the future reward is of equal importance as the current one \cite{barto2021reinforcement}. Most of the early works on average reward RL mainly focus on the tabular cases \cite{zhang2021average, zhang2021mean}, limiting their potential usage in complex environments. The initial development of average reward-based deep RL focused on Deep Q-Network (DQN) \cite{anschel2017averaged}, which has limited performance compared to the actor-critic-based methods. The recent advancement in actor-critic-based average reward DRL algorithms \cite{ma2021average, saxena2023off} has enabled the implementation of average reward RL for more practical problems.

\section{Problem Formulation} \label{sec:setting}

In this section, we first establish the RRM problem within the context of RAN slicing. Then, we formulate the RRM problem into a discounted reward RL one. We then discuss the mismatch between the objectives of these two formulations.

\subsection{RAN Slicing} \label{subsec:env}

\begin{figure}[hpbt]
    \centering
    \includegraphics[width = .8\columnwidth]{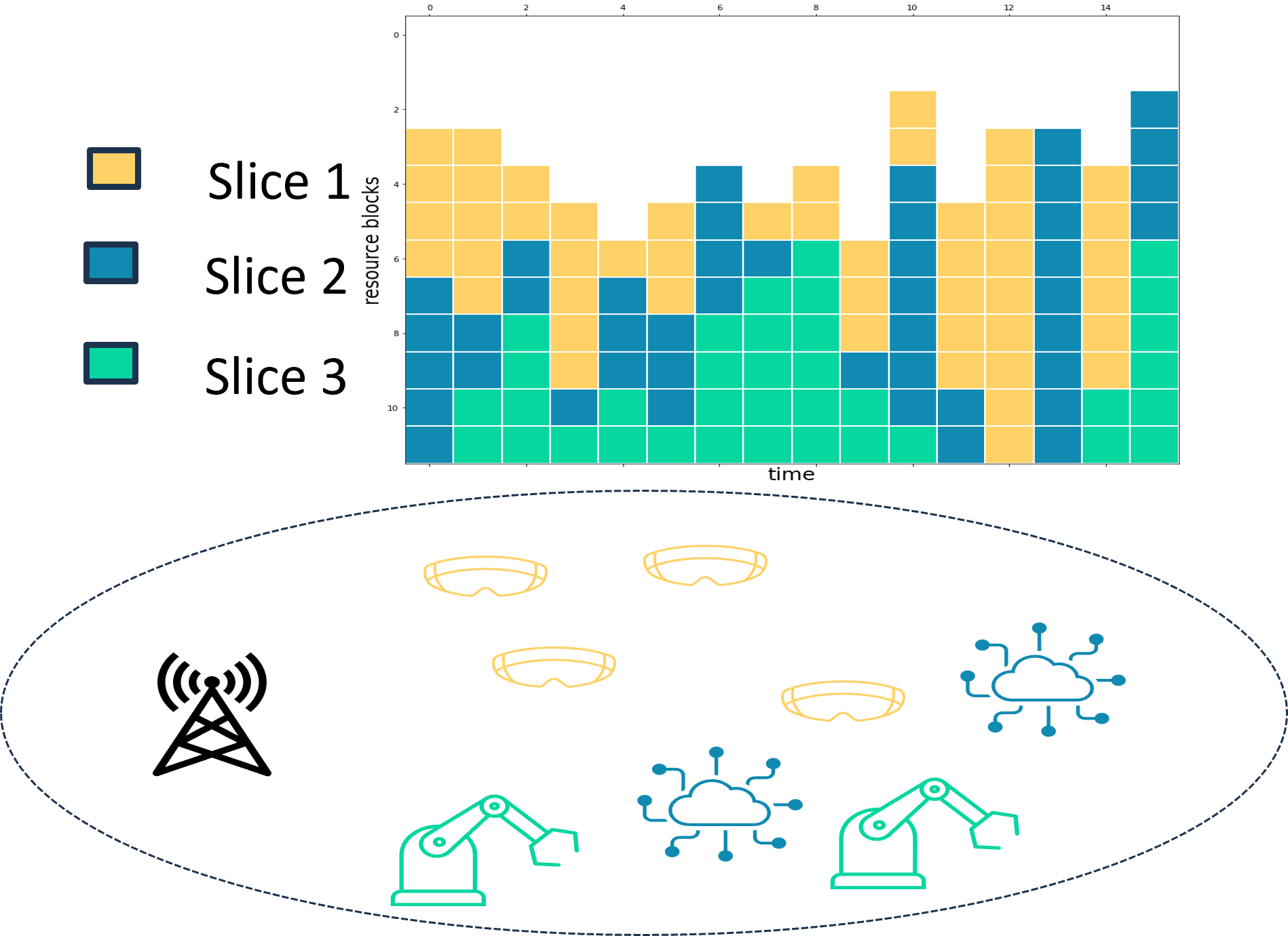}
    \caption{Illustration of a RAN slicing system}
    \label{fig:system}
\end{figure}

In a RAN slicing system, we assume that the system has $N$ slices in total, each handling a distinct user/traffic type. For these slices, our goal is to allocate packed radio resources properly to maximize the Quality of Service (QoS) of the whole system. These packed resources, named resource block groups (RBG), are then allocated to the users by the proportional fairness aware scheduler \cite{riley2010ns}. The system structure is illustrated in Fig. \ref{fig:system}. 

% \noindent
Assume the system has $M$ RBGs in total, and the QoS function at time $t$ is $f_t(\mathbf{M(t)})$, where $\mathbf{M(t)} = \left[ m_1(t), \cdots, m_N(t)  \right]$, where $\mathbf{M(t)}$ is the resource allocation vector which stands for the resource blocks allocated to different slices and $m_i(t)$ stands for the resource allocated to slice $i$. Then, we can formulate the optimization problem as:

\begin{equation} 
\label{eqn:rrmopt}
\begin{aligned}
& \underset{\mathbf{M(t)}}{\text{maximize}}
& & \underset{T \rightarrow \infty}{\text{lim}} \frac{1}{T} \sum_{t=1}^{T} f_t(\mathbf{M(t)}) \\
& \text{subject to}
& & \sum_{i = 1}^{N} m_i(t) \leq M. \
\end{aligned}
\end{equation}
% \congc{explain why the summation in the constraint above is only from 1 to $N-1$.}
Clearly, the design goal of this formulation is to achieve the best possible \emph{long-term average QoS}. 
% where it is obvious that the object is to reach a \textbf{maximum of average QoS return} over time.

\subsection{From RRM to Discounted Reward RL} \label{subsec:discounted}

In previous studies utilizing deep reinforcement learning (DRL) for the RRM problem in RAN slicing, the standard approach is to formulate the problem using discounted reward RL \cite{polese2022coloran, yang2023advancing, foukas2016flexran, Geng2017NetworkSlicing}. In this section, we first discuss this discounted reward RL setting and then show where the mismatch happens.

As a concrete case of RAN slicing, we consider that in the optimization problem outlined in Eq. \eqref{eqn:rrmopt}, two QoS metrics are pivotal: the total downlink throughput of the system and the average delay violation rate among users. The delay violation rate is defined as the proportion of packets exceeding the QoS latency threshold relative to the total number of packets a user receives. We then define the Markov Decision Process (MDP) for this RL problem as follows.

\begin{itemize}
    \item \textbf{Observations:} Building on the considerations outlined above regarding system performance metrics such as throughput and delay violation rate, it is pivotal to monitor how much of the allocated resources have been utilized. Accordingly, we gather the following metrics for each slice in the system to serve as observations in our MDP: received traffic throughput $T_{\text{rx}}$, traffic load $T_{\text{tx}}$, resource utilization rate $U$, delay violation rate $D_{\text{vio}}$, and average one-way delay $D_{\text{avg}}$ from every slice in the system. The observations are formally specified as
    \begin{equation*}
        \label{equ:state}
        \left\{ T_{\text{rx},i}, T_{\text{tx},i}, U_i, D_{\text{vio},i}, D_{\text{avg},i} \right \}_{i=1,\cdots, N}.
    \end{equation*}

    \item \textbf{Actions:} As outlined in Section \ref{subsec:env}, we need to allocate RBGs across different slices. Instead of distributing discrete resource units, our approach involves allocating a proportional share of RBGs to each slice, rendering our action a continuous variable within the range $[0, 1]$. Specifically, our action at time $t$ is represented as $A(t) = [a_{1}(t), \cdots, a_{N-1}(t)]$, where each $a_i(t) \in [0,1]$ denotes the proportion of RBGs allocated to slice $i$. We ensure the allocation is legitimate (i.e., $\sum_i a_i(t) \leq 1$) by integrating a softmax layer at the output of our policy network, ensuring a valid probability distribution over the slices.

    \item \textbf{Reward:} The reward design in a RAN slicing system should reflect its QoS objectives. Our configuration prioritizes two key components: the overall system throughput and delay violation rates. Accordingly, we construct our reward function as
    $$ R(t) = \sum_{i=1}^{N} r_i(t),$$ 
    where each component of the reward, $r_i(t)$, is defined as:     
    $$r_i(t) = T_{\text{rx}, i}(t) - \alpha  D_{\text{vio},i}(t).$$ 
    In our experiment, we set $\alpha = 4$  to impose a heavier penalty on the delay violations.
\end{itemize}

Assuming a discounted reward setting with the discount factor $\gamma$, the objective of this RL problem is:
$$
\underset{{\pi}}{\text{max}} \: \mathbb{E} \left[\sum_{t=0}^{\infty}\gamma^tR(t) \right].
$$

While this objective accumulates rewards over infinite time steps, \textit{the influence of future rewards diminishes significantly due to the discount factor}. For instance, with $\gamma = 0.95$, rewards beyond $50$ time steps contribute minimally to the objective, effectively accounting for only about $0.01$ of their original value. {This aspect of discounting does not align well with our initial goal as defined in Eq. \eqref{eqn:rrmopt}}, where the wireless network seeks optimal average performance over an infinite horizon. This mismatch motivates us to find better solutions to close the gap between the discounted RL and the original objective in our wireless network optimization problem.

\subsection{Detailed Environment Setting} \label{subsec:detail}

As described in Sec. \ref{subsec:env}, we consider an RRM problem in a RAN slicing system with $N$ slices and $M$ RBGs. In our  experiment, we have utilized \textbf{netgymenv} \cite{zhang2023netgym} as our simulator. We set $N=3$ and $M=25$. Our traffic model follows the LTE module in NS-3 \cite{riley2010ns}. To introduce different traffic flows for different slices, we assign a different user number to each of the slices ranging from $6$ to $20$. The detailed environment setting is given in Table \ref{tab:exp}. 

\begin{table}%[hpbt]
\caption{Experiment parameters}
\label{tab:exp}
\centering
\begin{tabular}{c|c}
\hline
Parameter & Value \\
\hline
Number of slices & 3 \\
Number of UEs per slice & $6-20$ \\
Delay violation threshold & 100 ms\\
Area & $120 \times 10$ m$^2$ \\
Downlink traffic & 2 Mbp/s\\
Traffic pattern & Poisson arrival\\
UE mobility & $1-2$ m/s\\
\hline
\end{tabular}
\end{table}

For the resource type we allocate to each of the slices, we utilize a setting similar to \cite{yang2023advancing} where the soft slicing strategy is used. In a soft slicing system, when the resource is allocated to a slice, the users in this slice have priority in using these resources. The leftover resources can then be re-used by other users from different slices if the allocated resource is not fully used. 

As for the RL algorithm, we use Soft Actor-Critic (SAC) \cite{haarnoja2018soft} as our primary choice. We choose this algorithm mainly because we would like to see whether extending an existing deep RL algorithm to its average reward version is applicable. 
% The training parameters of the neural network are listed in Table \ref{tab:train}. 

% \begin{table}%[hpbt]
% \caption{Training parameters}
% \label{tab:train}
% \centering
% \begin{tabular}{c|c}
% \hline
% Parameter & Value \\
% \hline
% Actor structure & 2 Layer MLP with hidden dimension 64\\
% Critic structure & 2 Layer MLP with hidden dimension 64\\
% Critic learning rate & $1e^{-4}$\\
% Actor learning rate & $3e^{-5}$\\
% $\rho$ learning rate* & $1e^{-5}$ \\
% \hline
% \end{tabular}
% \scriptsize
% \begin{center}
% *: only applicable to average reward RL models.
% \end{center}
% \end{table}

% \footnote{*: only applicable to average reward RL models. \congc{what is this??}}

\section{The impact of discount factor and horizon} \label{sec:understand}

We are not the first to identify the mismatch between the discounted reward RL and the real-world average return scenarios, where in \cite{zhang2020deeper}, the authors have noticed that there exists a $\gamma$ mismatch between the actors and critics. In \cite{tarasov2024revisiting, wu2022supported}, the authors report supreme performance with large $\gamma$ on long horizon tasks. In this section, we empirically establish that the same mismatch exists in the RRM problem for RAN slicing. 

We conduct two key experiments to validate the mismatch between the discounted reward RL objective and the real wireless system goal. To verify the impact of horizon length, we incorporate a period reset signal, which resets the simulator after a predefined time step $T$. We regard this reset length as the period length of our environment. In the first experiment, we fix $T$ and vary the discount factor $\gamma$, demonstrating that a larger $\gamma$ improves performance by valuing the longer future more equally. In the second experiment, we fix a large $\gamma$ and vary $T$, confirming that a larger $\gamma$ can help the agents look into the longer future.

\textbf{Fix $T$, vary $\gamma$:}  Table \ref{tab:vary_gamma} illustrates that when $T$ is constant, increasing $\gamma$ consistently enhances the RL agent's performance. The result suggests that it is helpful in a system trying to maximize long-term average rewards to have a larger discount factor, i.e. making the agent able to take longer steps into their consideration.

\begin{table}[hpbt]
    \centering
    \caption{Experimental results with $T = 200$ and different $\gamma$}
    \begin{tabular}{c|c}
    \hline
     $\gamma$  &  cumulative reward  \\
    \hline
       0.9     &  $10.53 \pm 1.25$   \\
    \hline
       0.93    &  $13.24 \pm 0.52$   \\
    \hline
       0.95    &  $14.20 \pm 0.50$   \\
    \hline
       \textbf{0.99} & $\mathbf{15.67 \pm 0.37}$\\
    \hline
    \end{tabular}    % \vspace{3mm}    
    \label{tab:vary_gamma}
\end{table}
\vspace{-3mm}

\textbf{Fix $\gamma$, vary $T$:} When $\gamma$ is fixed at a high value (e.g., 0.99), extending the horizon also results to an improved average reward per step, as evidenced by the results in Table \ref{tab:vary_t}. This longer horizon also plays a pivotal role as it ensures the RL agent can learn the transition from a longer future.

\begin{table}[hpbt]
    \centering
    \caption{Experimental results with $\gamma = 0.99$ and different $T$}
    \begin{tabular}{c|c}
    \hline
     $T$  &  average reward  \\
    \hline
       200     &  $0.078 \pm 0.002$   \\
    \hline
       500    &  $0.079 \pm 0.002$   \\
    \hline
       1000    &  $0.082 \pm 0.003$   \\
    \hline
       \textbf{2000} & $\mathbf{0.085 \pm 0.005}$\\
    \hline
    \end{tabular}
    % \vspace{-3mm}    
    \label{tab:vary_t}
\end{table}

\begin{figure}[hpbt]
    \centering
    \includegraphics[width = .8\columnwidth]{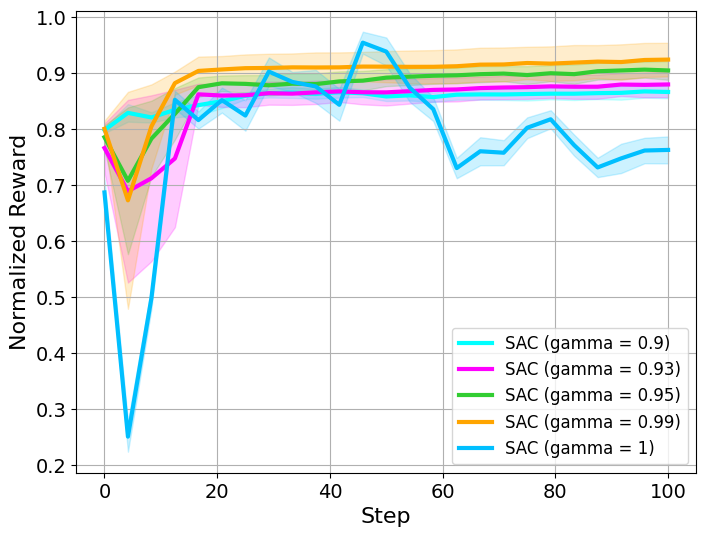}
    \caption{Experimental results with $\gamma = 1$. Shadowed areas indicate the confidence intervals.}
    \label{fig:gamma1}
\end{figure}

Summarizing these results, a heuristic solution emerges: set $\gamma=1$ which would ensure that rewards do not decrease over time. However, in our experiment shown in Figure \ref{fig:gamma1}, naively setting $\gamma$ to 1 appears beneficial for policy training initially but leads to significant instability later. This instability suggests that simply increasing $\gamma$ is not optimal. Based on this observation, a new tool is needed to close the gap between discounted reward RL and the network optimization goal.

\section{Average Reward Soft Actor-Critic} \label{sec:average}

The concept of average reward RL, as the name suggests, is to maximize the long-term average reward for sequential decision-making problems \cite{barto2021reinforcement}. In this section, we discuss the difference between the average reward RL and the discounted reward RL, and show how to change a discounted reward formulation into an average reward formulation. Then, following the design flow of ARO-DDPG \cite{saxena2023off}, we extend the SAC principle to satisfy the average reward objective and conduct experiments to evaluate the performance of this new algorithm.

\subsection{Re-formulation}

The difference between the average reward RL and discounted reward RL primarily lies in their objective functions, where the average reward RL's objective is defined as:
\begin{equation}
\label{eqn:avgrdRL}
\underset{{\pi}}{\text{max}} \: \underset{T \leftarrow \infty}{\text{lim}} \frac{1}{T} \sum_{t = 1}^{T} r(t).    
\end{equation}
In Eq.~\eqref{eqn:avgrdRL}, if we treat the reward function $r(t)$ as the QoS function in Eq. \eqref{eqn:rrmopt}, the goal of the average reward RL exactly matches the wireless network objective. %Which closed the gap we previously mentioned.

To solve this new RL problem, the major workflow remains the same. From the principle in \cite{barto2021reinforcement}, similar to the discounted reward setting, we can solve the average reward RL through the Bellman equation. More specifically, we can still utilize the TD error-based method to solve the problem. The only difference is that, instead of having a discount factor $\gamma$, we now maintain an estimation of the true average reward $\rho$ to help with the solution. The process of getting the average reward TD error is listed below.

% \congr{
In the average reward setting, we define the \textit{differential} return, which measures the difference between the rewards and the true average reward, as
\begin{equation}
    G_t \doteq R_{t+1} - r(\pi) +  R_{t+2} - r(\pi) +  R_{t+3} - r(\pi) + \cdots,
\end{equation}
where $\pi$ stands for the current policy. Based on this differential return, we then define the value function and compute the corresponding Bellman equation as
\begin{align}
    V_{\pi}(s) &= \mathbb{E}_{\pi}[G_t|S_t = s] \\
    & = \mathbb{E}_{\pi}[R_{t+1} - r(\pi) + G_{t+1}|S_t = s] \notag\\
    & = \underset{a}{\sum} \pi(a|s)\underset{s'}{\sum}\underset{r}{\sum} p(s', r| s, a)[r - r(\pi)   \notag\\
    &\quad+ \mathbb{E}_{\pi}[G_{t+1}|S_{t+1} = s']] \notag\\
    & = \underset{a}{\sum} \pi(a|s)\underset{r, s'}{\sum} p(s', r| s, a)[r - r(\pi) + V_{\pi}(s')].
\end{align}

From this Bellman equation and the definition of the TD difference, we can now compute the one-step TD as
\begin{equation}\label{equ:td-avg}
    \delta_t = R_{t+1} {- \rho} + V(S_{t+1}) - V(S_t). 
\end{equation}
For comparison, we note that the standard discounted reward TD error equals
\begin{equation} \label{equ:td-dis}
    \delta_t = R_{t+1} + {\gamma V(S_{t+1})} - V(S_t).
\end{equation}
Comparing these two, we see that the new TD error in Eq.~\eqref{equ:td-avg} substitutes the discount factor with the estimated average reward rate $\rho$.
% Compared to the discounted reward TD error in Eqn. \eqref{equ:td-dis}, this TD error substitutes the discount factor with the estimated average reward rate $\rho$.

% With the TD error, we can always use it to perform TD-style policy updates or deep RL algorithms following the same philosophy. 
% }

% \congc{you need to provide more details on how you get the new TD error. Cannot just directly give the result below.}
% \begin{align*}
% \textbf{Discounted Reward RL: }  & \delta_t = R_{t+1} + \gamma V(S_{t+1}) - V(S_t) \\
% \textbf{Average Reward RL: }  & \delta_t = R_{t+1} - \rho + V(S_{t+1}) - V(S_t). 
% \end{align*}
% Here, $R_{t+1}$ stands for the reward at time step $t+1$, $V(S_t)$ stands for the value function, and $\rho$ is the estimated average reward rate. 

\subsection{Average Reward SAC}

With the average reward TD error, if we have an accurate estimate of $\rho$, we can solve the RL problem by minimizing this error. However, this is a challenging task and 
%it is a hard topic to estimate this value under the deep learning setting accurately.
two main methods have been adopted in average reward DRL. One is to collect the full trajectory of the policy and directly estimate $\rho$ by setting $$\hat{\rho} = (1- \alpha)\hat{\rho} +  \frac{\alpha}{N} \sum_{n=1}^{N} r(s_n, a_n).$$ As described in \cite{ma2021average}, this type of estimation is more desirable for on-policy algorithms like PPO. The second choice is to make the average reward a trainable parameter and then to use gradient descent to update this parameter \cite{saxena2023off}. Mathematically we have $$ \hat{\rho}_{t + 1} = \hat{\rho}_{t} + \nabla_{\rho}\varepsilon_t,$$ where $$\varepsilon_t = r(s_t, a_t) - \rho_t - Q(s_t, a_t).$$ Since we use SAC as our primary discounted reward RL algorithm, which is an off-policy one, the latter choice is more applicable for designing the average reward SAC.

To develop SAC under an average reward setting, we take one step further from \cite{saxena2023off} and design $\varepsilon_t$ for SAC as: 
\begin{equation} \label{equ:average_td}
    \varepsilon_t = r(s_t, a_t) - \rho_t - \min(Q_1(s_t, a_t), Q_2(s_t, a_t)).
\end{equation}
Following this step, we extend the SAC algorithm into an average reward version and describe the complete procedure in Algorithm \ref{alg:ARO-sac}, where we mark the different steps incorporating $\rho$ in the {bold} font.

\begin{algorithm}
\caption{Average Reward Off-Policy Soft Actor-Critic (ARO-SAC)} \label{alg:ARO-sac}
\begin{algorithmic}[1] 
\State Initialize policy parameters $\theta$, Q-function parameters $\phi_1, \phi_2$, \textbf{average reward estimator $\rho$}
\State Initialize target Q-function parameters $\phi_{targ,1} = \phi_1, \phi_{targ,2} = \phi_2$
\State Initialize environment and observe initial state $s$
\State Initialize replay buffer $\mathcal{D}$

\For{each time step}
    \State Sample action $a \sim \pi_\theta(\cdot|s)$ based on current policy
    \State Execute action $a$ in the environment
    \State Observe reward $r$, new state $s'$, and done signal $d$
    \State Store transition tuple $(s, a, r, s', d)$ in replay buffer $\mathcal{D}$
    \State Sample random minibatch of transitions $(s, a, r, s', d)$ from $\mathcal{D}$
    \State \textbf{Compute target Q-value:}
    \State \hspace{\algorithmicindent} ${y = r - \rho + \min_{i=1,2} Q_{\phi_{targ,i}}(s', \tilde{a}')}$
    \State \hspace{\algorithmicindent} where $\tilde{a}' \sim \pi_\theta(\cdot|s')$
    \State Update Q-functions by one step of gradient descent using:
    \State \hspace{\algorithmicindent} $\nabla_{\phi_i} \frac{1}{|B|} \sum (Q_{\phi_i}(s, a) - y)^2$ for $i=1,2$
    \State Update policy by one step of gradient ascent using:
    \State \hspace{\algorithmicindent} $\nabla_\theta \frac{1}{|B|} \sum \log \pi_\theta(a|s) Q_{\phi}(s, a)$
    \State \textbf{Update average reward estimator} $\rho$:
    \State \hspace{\algorithmicindent} ${\nabla_\rho \frac{1}{|B|} \sum (\varepsilon_t)^2}$
    \State Update target networks:
    \State \hspace{\algorithmicindent} $\phi_{targ,i} \leftarrow \tau \phi_i + (1 - \tau) \phi_{targ,i}$ for $i=1,2$
    \State Observe new state $s \leftarrow s'$
\EndFor

\end{algorithmic}

\end{algorithm}

% where we marked the different steps incorporating $\rho$ in the \textbf{bold} font. %Notice that this algorithm is only an empirical extension without theoretical guarantee, which we might work on in our future research.

% And the new TD-error is defined as:

% \begin{equation}
%     \delta_t = R_{t + 1} - \rho + V(S_{t+1}) - V(S_t)
% \end{equation}

\subsection{Experiments} \label{subsec:exp}

\begin{figure}[hpbt]
    \centering
    \includegraphics[width = .8\columnwidth]{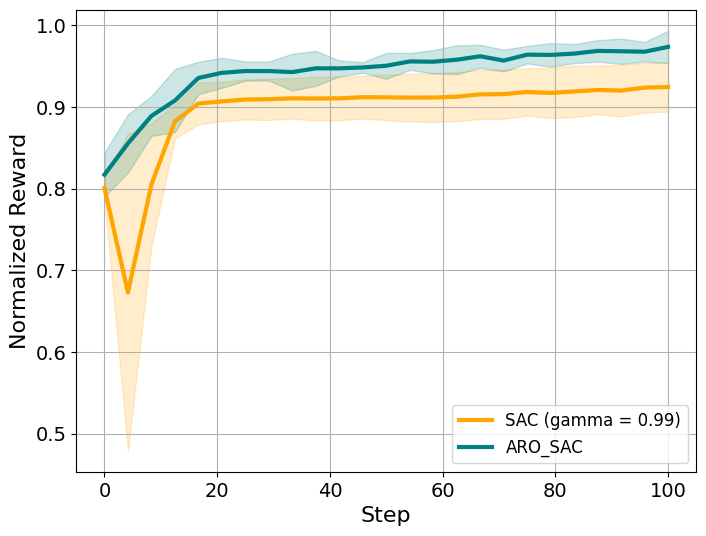}
    \caption{Experimental result using ARO-SAC, where the experiment is averaged over 5 independent runs over 5 different combinations of user numbers.}
    \label{fig:ARO-sac}
\end{figure}

In this section, we implement the proposed ARO-SAC in our simulation and compare its performance with the vanilla SAC with $\gamma = 0.99$. To verify the effectiveness of ARO-SAC, we compared the performance of the algorithm with pure SAC under horizon $T = 200$ and discount factor $\gamma = 0.99$.

The result in Fig. \ref{fig:ARO-sac} shows that the performance of our proposed ARO-SAC, while eliminating the drawback of an unstable convergence caused by setting $\gamma = 1$, also outperforms vanilla SAC with $\gamma = 0.99$ by $15 \%$. This demonstrates a solid gain of utilizing average reward RL on this RRM problem in RAN slicing over the discounted counterpart. However, we also want to point out that while the average reward RL does help with the policy's performance, it introduces an extra trainable parameter that needs extra hyperparameter tuning (learning rate selection) steps. The learning rate for the parameter $\rho$ needs careful selection. In our experiment, we set this learning rate to $1e-5$, which is slightly smaller than the learning rate of our actor-network.

\section{Conclusion} \label{sec:con}

This paper addressed a critical mismatch between the conventional discounted reward reinforcement learning (RL) framework and the long-term objectives inherent to radio resource management (RRM) in wireless networks. We first validated this mismatch between discounted reward objectives and the actual goals of wireless systems. Our results underscored that even slight modifications toward considering longer-term outcomes, such as extending the horizon and adjusting the discount factor, could enhance performance under the discounted reward framework. We then developed the Average Reward Off-policy Soft Actor-Critic (ARO-SAC), adapting the Soft Actor-Critic algorithm to the average reward framework, which significantly aligns with the long-term goals of RRM. Our experiments demonstrated a $15\%$ improvement in the overall system performance over the conventional discounted reward RL approach, confirming the effectiveness and advantages of average reward RL in enhancing wireless network management. Interesting future works include providing theoretical guarantees of ARO-SAC and improving the algorithm design by reducing the hyperparameter fine-tuning. 

% We also want to point out that although we extended SAC to ARO-SAC, this extension does not have solid theoretical backing. Designing the method with a theoretical guarantee could be a good direction for future research. The current average reward algorithm also needs an additional hyperparameter fine-tuning step, which introduces an extra workload. Investigating an algorithm without such restriction is also a potential direction we would like to investigate.  

% \section*{Acknowledgment}
% \section*{References}

\bibliographystyle{IEEEtran}
\bibliography{references}

\end{document}